\documentclass[twocolumn, twoside,prl]{revtex4}

\usepackage{amsmath}
\usepackage{amssymb}
\usepackage{latexsym}
\usepackage{revsymb}
\usepackage{graphicx}

\newcommand{\barr}{\begin{eqnarray}}
\newcommand{\earr}{\end{eqnarray}}
\newcommand{\ra}{\rangle}
\newcommand{\la}{\langle}
\newcommand{\beq}{\begin{equation}}
\newcommand{\eeq}{\end{equation}}

\begin{document}

\title{The quantum pigeonhole principle and the nature of quantum correlations}
\author{Y. Aharonov$^{1,2,3}$}
\author{F. Colombo$^{3,4}$}
\author{S. Popescu$^{3,5}$}
\author{I. Sabadini$^{3,4}$}
\author{D.C.Struppa$^{2,3}$}
\author{J. Tollaksen$^{2,3}$}
\affiliation{$^1$ School of Physics and Astronomy, Tel Aviv University,  %
Tel Aviv, Israel}
\affiliation{$^2$ Schmid College for Science and Technology, Chapman University, 1 University Dr, Orange, CA, USA}
\affiliation{$^3$ Institute for Quantum Studies, Chapman University, 1 University Dr, Orange, CA, USA}

\affiliation{$^4$ Politecnico di Milano, Dipartimento di Matematica, Via E. Bonardi, 9 20133 Milano, Italy}
\affiliation{$^5$ H.H.Wills Physics Laboratory, University of Bristol, %
 Tyndall Avenue, Bristol BS8 1TL, U.K.}

\begin{abstract}

\end{abstract}

\date{}

\maketitle

\bigskip
\noindent
{\bf The pigeonhole principle: "If you put three pigeons in two pigeonholes at least two of the pigeons end up in the same hole" is an obvious yet fundamental principle of Nature as it captures the very essence of counting. Here however we show that in quantum mechanics this is not true! We find instances when three quantum particles are put in two boxes, yet no two particles are in the same box. Furthermore, we show that the above "quantum pigeonhole principle" is only one of a host of related quantum effects, and points to a very interesting structure of quantum mechanics that was hitherto unnoticed. Our results shed new light on the very notions of separability and correlations in quantum mechanics and on the nature of interactions. It also presents a new role for entanglement, complementary to the usual one. Finally, interferometric experiments that illustrate our effects are proposed.}

\section{The quantum pigeonhole principle}
\bigskip
\noindent
Arguably the most important lesson of quantum mechanics is that we need to critically revisit our most basic assumptions about Nature. It all started with challenging the idea that particles can have, at the same time, both a well-defined position and a well-defined momentum, and went on and on to similar paradoxical facts. But the pigeon-hole principle that is the subject of our paper seems far less likely to be challenged. Indeed, while on one hand it relates to physical properties of objects - it deals, say, with actual pigeons and pigeon-holes - it also encapsulates abstract mathematical notions that go to the core of what numbers and counting are so it underlies, implicitly or explicitly, virtually the whole of mathematics. (In its explicit form the principle was first stated by Dirichlet in 1834 \cite{dirichlet} and even in its simplest form its uses in mathematics are numerous and highly non-trivial \cite{math-uses}.)  It seems therefore to be an abstract and immutable truth, beyond any doubt. Yet, as we show here, for quantum particles the principle does not hold.

\bigskip
\noindent
Consider three particles and two boxes, denoted $L$ (left) and $R$ (right).
To start our experiment, we prepare each particle in a superposition of being in the two boxes, \beq |+\ra={1\over{\sqrt2}}\Big( |L\ra +|R\ra\Big).\eeq The overall state of the three particles is therefore

\beq|\Psi\ra=|+\ra_1|+\ra_2|+\ra_3.\label{preselected}\eeq

\bigskip
\noindent
Now, it is obvious that in this state any two particles have non-zero probability to be found in the same box. We want however to show that there are instances in which we can guarantee that no two particles are together; we cannot arrange that to happen in every instance, but, crucially, there are instances like that. To find those instances we subject each particle to a second measurement: we measure whether each particle is in the state $|+i\ra={1\over{\sqrt2}}\Big( |L\ra+i|R\ra\Big)$ or $|-i\ra={1\over{\sqrt2}}\Big( |L\ra -i|R\ra\Big)$ (these are two orthogonal states, so there is an operator whose eigenstates they are - we measure that operator). The cases we are interested in are those in which all particles are found in $|+i\ra$, i.e. the final state

\beq|\Phi\ra=|+i\ra_1|+i\ra_2|+i\ra_3.\label{postselected}\eeq

\bigskip
\noindent Importantly, neither the initial state nor the finally selected state contain any correlations between the position of the particles. Furthermore, both the preparation and the post-selection are done independently, acting on each particle separately.

\bigskip
\noindent
Let us now check whether two of the particles are in the same box. Since the state is symmetric, we could focus on particles 1 and 2 without any loss of generality - any result obtained for this pair applies to every other pair.

\bigskip
\noindent
Particles 1 and 2 being in the same box means the state being in the subspace spanned by $|L\ra_1|L\ra_2$ and $|R\ra_1|R\ra_2$; being in different boxes corresponds to the complementary subspace, spanned by $|L\ra_1|R\ra_2$ and $|R\ra_1|L\ra_2$. The projectors corresponding to these subspaces are
\barr \Pi^{same}_{1,2}&=& \Pi^{LL}_{1,2}+\Pi^{RR}_{1,2}\nonumber\\ \Pi^{diff}_{1,2}&=& \Pi^{LR}_{1,2}+\Pi^{RL}_{1,2}\earr
where

\barr\Pi^{LL}_{1,2}&=&|L\ra_1|L\ra_2{\,}_1\la L|_2\la L|,~~\Pi^{RR}_{1,2}=|R\ra_1|R\ra_2{\,}_1\la R|_2\la R|,~~~~~~~\nonumber\\
\Pi^{LR}_{1,2}&=&|L\ra_1|R\ra_2{\,}_1\la R|_2\la L|,~~\Pi^{RL}_{1,2}=|R\ra_1|L\ra_2{\,}_1\la L|_2\la R|.\earr

\bigskip
\noindent
On the initial state alone, the probabilities to find particles 1 and 2 in the same box and in different boxes are both $50\%$. On the other hand, given the results of the final measurements, we always find particles 1 and 2 in different boxes. Indeed, suppose that at the intermediate time we find the particles in the same box. The wavefunction then collapses (up to normalisation) to

\beq |\Psi'\ra=\Pi^{same}_{1,2}|\Psi\ra={1\over2}\big(|L\ra_1|L\ra_2+|R\ra_1|R\ra_2\big)|+\ra_3\eeq
which is orthogonal to the post-selected state (\ref{postselected}), i.e.

\beq \la\Phi| \Pi^{same}_{1,2}|\Psi\ra=0\label{pigeonhole}.\eeq
Hence in this case the final measurements cannot find the particles in the state $|\Phi\ra$. Therefore the only cases in which the final measurement can find the particles in the state $|\Phi\ra$ are those in which the intermediate measurement found that particles 1 and 2 are in different boxes.

\bigskip
\noindent Crucially, as noted before, the state is symmetric under permutation, hence what is true for particles 1 and 2 is true for all pairs. In other words, given the above pre- and post-selection, we have three particles in two boxes, yet no two particles can be found in the same box - our quantum pigeonhole principle.

\section{A related effect for every final outcome}

\bigskip
\noindent
In the previous section we focused on what happens when at the final measurement all particles are found in the state $|+i\ra$. It is in this case that the intermediate measurements exhibit the quantum pigeon-hole effect. However this is only one of the eight possible outcomes of the final measurement - indeed, each particle can be found either in $|+i\ra$ or in $|-i\ra$. It is important to note that in each of these cases an interesting effect occurs: In the case when the final state is $|-i\ra_1|-i\ra_2|-i\ra_3$ the intermediate measurements exhibit once again the quantum pigeon-hole effect, i.e. no two particles can be found in the same box. For the final state $|-i\ra_1|+i\ra_2|+i\ra_3$ intermediate measurements find that particle 2 is in the same box as 1, particle 3 is in the same box as 1 but particles 2 and 3 are not in the same box (see supplementary information). Similar patterns occur in all other cases.

\section{Generalizing the quantum pigeonhole principle}

\bigskip
\noindent
The above effect is but one of a multitude of similar effects. For example, in the case of $N$ particles in $M<N$ boxes we can guarantee that no two particles are in the same box when we prepare each particle in the state

\beq |0\ra={1\over{\sqrt M}}\sum_{k=1}^M |k\ra\eeq
and we find, in a final measurement, each particle in the state
\beq |{{\pi}\over M}\ra={1\over{\sqrt M}}\sum_{k=1}^M e^{i{{\pi k}\over M}}|k\ra\eeq
(see supplementary information).

\section{The nature of quantum correlations}

\noindent
Before analyzing our paradox in more detail, we would like to comment more on the nature of quantum correlations.

\bigskip
\noindent The first thing to notice is that neither the pre-selected state nor the post-selected state are correlated (they are both direct products and each particle is prepared and post-selected individually) yet the particles are correlated.

\bigskip
\noindent The second thing to notice is that if we measure the location of each particle individually, they appear to be completely uncorrelated. Indeed, suppose we measure separately the location of particle 1 and 2. There are four possible outcomes of this measurement: $LL$, $LR$, $RL$ $RR$ and, as one can easily show,  they all occur with equal probabilities.  It is only when we ask {\it solely about the correlation, and no other information}, (i.e whether the two particles are in the same box or not, without asking in which box they are), that we find them correlated.

\bigskip
\noindent The above shows a fundamental difference in the way in which the probabilities work in the standard, "pre-selected only" experiment and in a "pre- and post-selected" one (i.e. when we only consider the cases in which a final measurement gave a particular answer).

\bigskip
\noindent Indeed, consider first the standard situation, that is, consider that the particles are prepared in the state $|\Psi\ra$ and we do not perform a final measurement and selection according to its result. Suppose first that we measure separately the location of each particle. The probabilities to find $LL$ and $RR$ are given, respectively by

\beq P(LL)=\la\Psi|\Pi^{LL}_{1,2}|\Psi\ra~~~~P(RR)=\la\Psi|\Pi^{RR}_{1,2}|\Psi\ra.\eeq
Hence, the probability to find the particles in the same box by using this measurement is
\barr &&P(LL~{\rm or}~RR)= P(LL)+P(RR)=~~~~~~~~~~~~~~~~~~~~~~~~\nonumber\\&& \la\Psi|\Pi^{LL}_{1,2}|\Psi\ra+\la\Psi|\Pi^{RR}_{1,2}|\Psi\ra=\la\Psi|\Pi^{LL}_{1,2}+\Pi^{RR}_{1,2}|\Psi\ra.\earr

\bigskip
\noindent
Suppose however that we measure an operator that only tells us whether or not the particles are in the same box, without indicating in which box they are. This operator has only two eigenvalues, corresponding to "same" and "different" and the corresponding projectors
$\Pi^{same}_{1,2}$ and $\Pi^{diff}_{1,2}$. The probability to find the two particles in the same box by this measurement is then
\beq P(same)=\la\Psi|\Pi^{same}_{1,2}|\Psi\ra=\la\Psi|\Pi^{LL}_{1,2}+\Pi^{LL}_{1,2}|\Psi\ra\eeq
which is identical to the probability of finding the particles in the same box when we measure their position individually. Hence for a standard, pre-selected only situation, the course-grained measurement that asks only about correlations but no other information gives the same probabilities as the more detailed measurement.

\bigskip
\noindent
On the other hand, suppose we compare the above two measurement methods but in the case of a pre- and post-selected ensemble. In full generality, when one measures an arbitrary operator $A$ on a pre and post selected ensemble, the probability to obtain the eigenvalue $a_i$ is given by
\beq P(a_i)={{\Big|\la\Phi|\Pi_i|\Psi\ra\Big|^2}\over{\sum_j\Big|\la\Phi|\Pi_j|\Psi\ra\Big|^2}}\eeq
where $\Pi_i$ are the corresponding projectors \cite{abl}.

Using this result, in the case of separate measurements on each particle we find

\beq P(LL)={1\over N}\Big|\la\Phi|\Pi^{LL}_{1,2}|\Psi\ra\Big|^2,~~~P(RR)={1\over N}\Big|\la\Phi|\Pi^{RR}_{1,2}|\Psi\ra\Big|^2\eeq
with
\beq N=\sum_{kl=L,R}\Big|\la\Phi|\Pi^{kl}_{1,2}|\Psi\ra\Big|^2\eeq
Obviously $P(LL~{\rm or}~RR)= P(LL)+P(RR)$ is now different from
\beq P(same)={{\Big|\la\Phi|\Pi^{same}_{1,2}|\Psi\ra\Big|^2}\over{\Big|\la\Phi|\Pi^{same}_{1,2}|\Psi\ra\Big|^2+
\Big|\la\Phi|\Pi^{diff}_{1,2}|\Psi\ra\Big|^2}}.\eeq

\bigskip
\noindent
What the above discussion shows is that there is a significant difference between correlations that can be observed when we measure particles separately and when we measure them jointly.
This difference can be observed only when we consider pre- and post-selected ensembles, but it is always there, as an intrinsic part of quantum mechanics. Indeed, one may not be familiar with the idea of pre- and post-selection but it fact it is something that we encounter routinely:  Every time when we have a sequence of measurements we can split the original ensemble into a number of different pre and post-selected sub-ensembles according to the result of the final measurement, and in each such sub-ensemble we can observe a similar effect.

\bigskip
\noindent The third thing to notice is that the global measurement which only asks about correlations but no other detailed information is, in some sense, better than the detailed measurement as it delivers the information about correlations while minimizing the disturbance that it produces to the state. Indeed, suppose two particles are in an arbitrary superposition $\alpha |L\ra_1|R\ra_2+\beta|R\ra_1|L\ra_2$. The global measurement will tell us that the particles are in different boxes and will not disturb the state at all, since it is an eigenstate of the measured operator. On the other hand, if we measure each particle separately, we disturb the state, collapsing it on either $|L\ra_1|R\ra_2$ or $|R\ra_1|L\ra_2$. In the case of a pre-selected only ensemble, what happens to the state after the measurement doesn't matter, but if we are interested in following this measurement with a subsequent measurement and look at the different pre- and  post-selected ensembles, how much the intermediate time measurement disturbed the state is essential. This is the reason why post-selection is essential in order to see the effect.

\bigskip
\noindent Finally, and most importantly, we note that the global measurement is a measurement of an operator with entangled eigenstates and it requires either to put the particles in interaction or consume some entanglement resources to perform it. The quantum pigeonhole effect is thus an example of a new aspect of entanglement: entanglement in the measurement is needed to reveal correlations existing in a direct product state.

\section{The nature of quantum interactions. A first experiment.}

\bigskip
\noindent
The quantum pigeonhole effect has major implications for the understanding of the very nature of quantum interactions. Consider again three particles and two boxes. Let the particles interact with each other by bipartite short-range interactions, i.e. any two particles interact when they are in the same box and do not interact otherwise. Then, as there are three particles and only two boxes we expect that {\it always} at least two of the particles should interact. But, due to our pigeonhole effect, this is not so, as shown in the following experiments.

\bigskip
\noindent
Consider a Mach-Zender interferometer for electrons, as depicted in Fig. 1. It consists of two beam-splitters BS1 and BS2 a phase shifter PS that introduces a phase-shift of $\pi$ and two detectors, $D_1$ and $D_2$. The detectors have spatial resolution, so they can tell exactly where each particle landed. When an electron is injected from the left side, the first beam-splitter generates the state $|+\ra={1\over{\sqrt2}}\Big( |L\ra +|R\ra\Big)$. The phase shifter, final beam-splitter and the detectors in effect implement a measurement with eigenstates $|\pm i\ra={1\over{\sqrt2}}\Big( |L\ra \pm|R\ra\Big)$. Indeed, if the state before PS is $|+i\ra$ the electron ends, with certainty at $D_1$ while if it is $|-i\ra$ it ends with certainty at $D_2$.

\bigskip
\noindent
Suppose now that we inject simultaneously three electrons in the interferometer from the left side, such that they travel in parallel beams. The beams are arranged, as shown in Fig 1a, in an equilateral triangle configuration.

\bigskip
\noindent When two electrons go through the same arm of the interferometer they repel each other and their trajectories are deflected. Indeed, the force that one electron exerts on the other produces a change in momentum and this in turn leads to the deflection of the beams by an amount depending on the original separation of the beams, the length of the interferometer and the speed of the electrons.  When the electrons go through different arms they effectively do not interact (since the arms are separated by a large distance). Since we have three electrons and only two arms, we expect to always have interactions, regardless of which detectors the electrons end up at after traversing the interferometer.

\begin{figure}
\includegraphics[width=0.6\columnwidth]{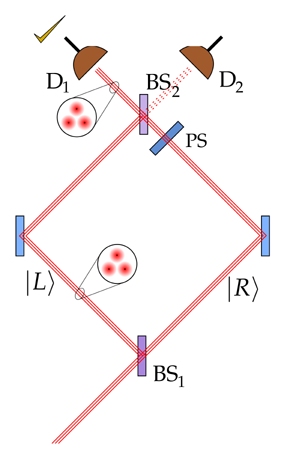} \caption{A Mach-Zender interferometer for electrons. $BS_1$ and $BS_2$ are beam-splitters, $PS$ is a $\pi$ phase shifter and $D_1$ and $D_2$ are detectors with spatial resolution. Three electrons are simultaneously injected on parallel trajectories. The cross section shows the spatial arrangement of the beams. The image at the detector is obtained by superposing the results of many runs, but only keeping those runs in which all three electrons arrived at $D_1$.}
\end{figure}

\begin{figure}

\includegraphics[width=0.8\columnwidth]{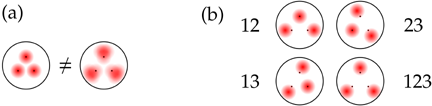} \caption{ (a) The observed distribution shows no displacement of the beams, while naively we would expect the beam to be moved and deformed, due to to a combination of different cases depicted in (b), which depend on which particles are in the same arm and repel each other.}

\end{figure}

\bigskip
\noindent We are interested in what happens in the cases when all three electrons end up at $D_1$ and when the interaction between the electrons is not too strong. Ending up at $D_1$ is our desired post-selection. The requirement that the interaction should not be too strong (which can be arranged, for example, by spacing the parallel beams further away or making the arms shorter) is more subtle. The point is the following. We know that when considering three non-interacting particles in two boxes, given the appropriate pre and post-selection, no two particles are in the same box. We certified this by showing that whichever pair of particles we measured we always find them in different boxes. But if we were to try to measure two or all pairs in the same experiment, the measurements would disturb one another and we would not see the effect. Indeed, suppose we first measure if particles 1 and 2 are in the same box, then particles 1 and 3 and then we make the post-selection. Now it is actually possible to find 1 and 2, or 1 and 3 or both 1 and 2 and 1 and 3 in the same box. Indeed, for example starting with the initial state (\ref{preselected}) we can find particles 1 and 2 in the same box. In this case the state collapses to $|\Psi'\ra=\Pi^{same}_{1,2}|\Psi\ra$. This state is orthogonal to the post-selected state (\ref{postselected}), so if we were to make the post-selection now this even would not be selected - this was our main proof of the quantum pigeon-hole effect. However, suppose that instead of making the post-selection at this point we now make a measurement of the pair 1,3. We can also find 1 and 3 in the same box. The state now collapses to $|\Psi''\ra=\Pi^{same}_{1,3}\Pi^{same}_{1,2}|\Psi\ra$. But now, crucially, $|\Psi''\ra$ is not orthogonal to the post-selected state (\ref{postselected}) hence these events are not eliminated from the post-selected ensemble. To be able to see the effect we need to limit the mutual disturbance of the measurements.

\bigskip
\noindent
In our interferometer case, the interaction between the electrons leads to the deflection of the beams whenever two electrons are in the same arm. The deflection of the beams is therefore effectively like a measurement of whether or not two electrons are in the same arm. To limit the disturbance produced by the simultaneous existence of the interaction between all three pairs, we just need to reduce the strength of the effect of the interaction. Technically, we want to ensure that the change in the momentum of an electron due to the force produced by the other electrons is smaller than the spread in its momentum. Due to this, the deflection of a beam - if it occurs - is smaller than it's spatial spread, hence, by seing where one electron landed on $D_1$ we cannot tell if its trajectory was deflected or not. However, by collecting the results of multiple runs we can easily identify if the electrons were deflected or not (see Fig. 2).

\bigskip
\noindent Naively, we would expect to see the three beams deflected outwards and deformed, - each electron should move radially outwards when all three are present in the same arm, and sideways when only two electrons are present. We expect the deviation to be by less than the cross section but nevertheless by a noticeable amount. Instead (see supplementary information) what we find is that the beams are completely un-deflected and undisturbed (up to second order perturbations), indicating that indeed there was no interaction whatsoever between the electrons.

\section{A second experiment}

\bigskip
\noindent
A second experiment uses a similar interferometer as described in the first experiment, but instead of electrons we now inject atoms. Let all the atoms start in an excited state and arrange the setting in such way that there is a very significant probability for the atoms to spontaneously emit photons while traversing the interferometer. We surround the interferometer with photon detectors that could detect the emitted photons. Importantly, we chose the energy separation between the ground and excited state such that the wave-length of the emitted photons is much larger than the separation between the arms,so that by detecting a photon we cannot tell whether the atom that emitted it went through the left or right arm. Again, we inject all three atoms from the left and are interested only in the cases in which all three end up at detector $D_1$.

\bigskip
\noindent When two atoms are close to each other (being in the same arm) they interact with each other (for example by dipole-dipole interactions) and the energy levels are shifted. Observing the wavelength of the emitted photons we can then tell if the atoms were in the same arm or not. Following the same reasoning as used in the previous experiment, we also arrange that the frequency shift is smaller than the spread of the spectral line, so that the three pairwise interactions should not disturb each other. Again, due to this, one single photon observation cannot tell us whether the frequency was shifted or not, but accumulating the statistics we can detect the shift.

\bigskip
\noindent Since there are three atoms but only two arms, similarly to the previous experiment, we expect that in each run of the experiment at least two atoms will be in the same arm, so the photons they emit will be frequency shifted, no matter which detector the atoms end up in. Yet, according to our quantum pigeonhole effect, when we look at the cases in which all three atoms end up at $D_1$, we see that the spectral lines are unshifted.

\section{Conclusions}

\bigskip
\noindent
In conclusion we presented a new quantum effect that requires us to revisit some of the most basic notions of quantum physics - the notions of separability, of correlations and of interactions. It is still very early to say what the implications of this revision are, but we feel one should expect them to be major since we are dealing with such fundamental concepts.

\section{Supplementary information}

\section {Generalizing the pigeonhole effect}

\bigskip
\noindent
Consider $N$ particles in $M<N$ boxes. Let the initial state of each particle be
\beq |0\ra={1\over{\sqrt M}}\sum_{k=1}^M |k\ra.\eeq
Suppose that finally we perform on each particle a measurement of an operator that has the state
\beq |{{\pi}\over M}\ra={1\over{\sqrt M}}\sum_{k=1}^M e^{i{{\pi k}\over M}}|k\ra\eeq
as one of its eigenstates, and suppose we find each particle in this state. In other words, let the initial state be
\beq |\Psi\ra=|0\ra_1|0\ra_2...|0\ra_N\eeq
and the final measurement finds the overall state
\beq |\Phi\ra=|{{\pi}\over M}\ra_1|{{\pi}\over M}\ra_2...|{{\pi}\over M}\ra_N.\eeq
Suppose now that at the intermediate time we measure whether or not particles 1 and 2 are in the same box. The projector corresponding to the particles being in the same box is
\beq \Pi^{same}_{1,2}=\sum_{j=1}^M|j\ra_1|j\ra_2{}_2\la j|_1\la j|.\eeq
Suppose now that at the intermediate measurement we do find particles 1 and 2 in the same box. The initial state then collapses to $\Pi^{same}_{1,2}|\Psi\ra$ which is orthogonal to $|\Phi\ra$. Indeed
\beq \la\Phi|\Pi^{same}_{1,2}|\Psi\ra=\sum_{k=1}^M {{2\pi k}\over M}=0.\eeq
Therefore if at the intermediate time we do find particles 1 and 2 in the same box, we couldn't find $|\Phi\ra$ at the final measurement, which means that if we did find $|\Phi\ra$ at the final measurement, the intermediate measurement found particles 1 and 2 to be in different boxes. Again, since both the initial and the final states are symmetric under permutation, the same conclusion holds for any pair of particles. In other words, we have a situation where $N$ particles are in $M<N$ boxes, but no two particles are in the same box.

\bigskip
\section{The interferometric experiment}

\bigskip
\noindent
To describe our interferometric experiment it is useful to separate he degrees of freedom into a "which arm" degree of freedom and the relative positions $r_{ij}$ of the electrons when they are in the same arm. Using this decomposition, the initial wavefunction inside the interferometer is $|\Psi\ra|\varphi\ra$ where $|\Psi\ra=|+\ra_1|+\ra_2|+\ra_3$ and $|\varphi\ra$ describes the relative degrees of freedom. The interaction Hamiltonian between the electrons can be approximated by
\beq H_{int}=\sum_{i,j=1}^3\Pi^{same}_{i,j}V(r_{i,j})\eeq
where $V$ is the interaction potential.  The projector $\Pi^{same}_{i,j}$ describes the fact
that the electrons interact only when they are in the same arm. Furthermore, the "which arm" degree of freedom is a constant of motion (i.e. the electrons do not go from one arm to the other).  The time evolution of the state, in the interaction picture is then given by
\beq e^{-iH_{int}T}|\Psi\ra|\varphi\ra\eeq where $T$ is the total time of the interaction (the time taken for the electrons to go through the interferometer) and where for simplicity, we set $\hbar=1$. We are interested in the relative degrees of freedom (the detailed positions of the beams) given the post-selection, i.e. in
\beq \la\Phi|e^{-iH_{int}T}|\Psi\ra|\varphi\ra\eeq
where $|\Phi\ra=|+i\ra_1|+i\ra_2|+i\ra_3$ is the post-selected state.
Given that we take the interaction to be relatively weak, we could approximate it by the first order perturbation
\beq \la\Phi|(1-iH_{int}T)|\Psi\ra|\varphi\ra=\la\Phi|\Psi\ra|\varphi\ra - iT\la\Phi|H_{int}|\Psi\ra|\varphi\ra.\eeq
But
\barr &&\la\Phi|H_{int}|\Psi\ra=\la\Phi|\sum_{i,j=1}^3\Pi^{same}_{i,j}V(r_{i,j})|\Psi\ra\nonumber\\&&
=\sum_{i,j=1}^3\la\Phi|\Pi^{same}_{i,j}|\Psi\ra V(r_{i,j})=0\earr
where the last equality follows from the fact that $\la\Phi|\Pi^{same}_{i,j}|\Psi\ra=0$  which is the basis for the pigeonh0le effect. Hence, in the first order of approximation, given the success of the post-selection, there is effectively no interaction between the electrons.

\end{document}